\documentclass[manuscript]{aastex}

\usepackage{color}
\usepackage{soul}

\shorttitle{New spectral characterization of new experimental measurements of dimethyl ether isotopologues}
\shortauthors{Fern\'andez et al.}

\begin{document}
\title{New spectral characterization of dimethyl ether isotopologues CH$_3$OCH$_3$ and $^{13}$CH$_3$OCH$_3$ in the THz region}

\author{J.M. Fern\'andez and G. Tejeda
}
\affil{Laboratory of Molecular Fluid Dynamics, Instituto de Estructura de la Materia IEM-CSIC, Unidad Asociada GIFMAN, CSIC-UHU; 28006 Madrid, Spain; }
\email{jm.fernandez@csic.es}

\and

\author{M. Carvajal}
\affil{Dpto. Ciencias Integradas, Centro de Estudios Avanzados en
  F\'{\i}sica, Matem\'atica y Computaci\'on, Facultad de Ciencias Experimentales, Universidad de Huelva; Unidad Asociada GIFMAN, CSIC-UHU; 21071 Huelva, Spain}

\affil{Instituto Universitario Carlos I de F\'{\i}sica Te\'orica y
  Computacional, Universidad de Granada, Granada, Spain}

\and

\author{M.L. Senent}
\affil{Theoretical Chemistry and Physics Department, Instituto de Estructura de la Materia IEM-CSIC, Unidad Asociada GIFMAN, CSIC-UHU; 28006 Madrid, Spain; }

\begin{abstract}
The torsional Raman spectra of two astrophysically detected isotopologues of
dimethyl-ether, ($^{12}$CH$_3$O$^{12}$CH$_3$ and
$^{13}$CH$_3$O$^{12}$CH$_3$), have been recorded at room temperature and
cooled in supersonic jet, and interpreted with the help of highly correlated ab initio calculations.
Dimethyl-ether displays excited torsional and vibrational levels at low energy that can be populated at the temperatures of
the star forming regions, obliging to extend the analysis of the rotational spectrum over
the ground state. Its spectrum in the THz region 
is rather complex due to the coupling of the torsional overtones
$2\nu_{11}$ and $2\nu_{15}$ with the COC bending mode, and
the presence of many hot bands. 
The torsional overtones are set here at
$2\nu_{11}=385.2$~cm$^{-1}$ and $2\nu_{15}=482.0$~cm$^{-1}$
for $^{12}$CH$_3$O$^{12}$CH$_3$,  and 
$2\nu_{11}=385.0$~cm$^{-1}$ and $2\nu_{15}=481.1$~cm$^{-1}$
for $^{13}$CH$_3$O$^{12}$CH$_3$.
The new assignment of $2\nu_{11}$ is
downshifted around $\sim 10$~cm$^{-1}$ with respect to the literature.
All the other (hot) bands have been re-assigned consistently.
In addition, the infrared-forbidden torsional fundamental band $\nu_{11}$
is observed here at 197.8~cm$^{-1}$. The new spectral characterization
in the THz region reported here provides improved values of the
Hamiltonian parameters, to be used in the analysis of the rotational
spectra of DME isotopologues for further astrophysical detections.
\end{abstract}

\newpage

\section{Introduction}

Dimethyl ether (DME, CH$_3$OCH$_3$) is a relevant
astrophysical molecule which was first detected in the ISM by~\citet{snyder1974},
and later on, identified as an abundant species in star forming regions~\citep{schilke2001}.
DME is mainly formed in the gas phase via the radiative
association reaction of methoxy and methyl radicals, and it presents a
correlation with methyl formate~\citep{carvajal2010,favre2014},
via the oxidation of CH$_3$OCH$_2$, in cold objects ~\citep{balucani2015}.  

In addition to works prior to its detection in the ISM
~\citep{taylor1957,kasai1959,fateley1962,blukis1963,durig1976,groner1977},
the astrophysical interest for the complete description of the
spectrum in the millimeter and submillimeter regions of 
DME and its isotopologues, such as $^{13}$CH$_3$OCH$_3$ ($^{13}$C-DME) and
CH$_3$OCH$_2$D (d-DME), has brought about
many laboratory studies of its rotational spectrum and of a few vibrational bands
~\citep{lovas1979,neustock1990,groner1998,coudert2002,niide2003,endres2009,kutzer2016}. 
In turn, the analysis of an extensive number of spectral lines in astronomical observations
has given rise to the identification of DME in Orion-KL ~\citep{brouillet2013} and the
detection of the first excited torsional states lines in a high-mass
star forming region ~\citep{bisschop2013}.
Furthermore, the ground state rotational lines corresponding to 
the monosubstituted isotopologues $^{13}$C-DME and d-DME were first
observed in a high-mass star forming region ~\citep{koerber2013,richard2013}.

DME is a non-rigid asymmetric-top with two methyl internal
rotors that splits each rovibrational level into nine components
~\citep{groner1977,senent1995a,senent1995b}, some of them degenerate.
Hence, the high-resolution spectrum of DME is quite dense and its
analysis challenging. Furthermore, because the first and the second
excited torsional levels lie at relatively low energy, they can be
populated at the temperatures of the hot core regions obliging to
extend the spectral analysis to the two torsional fundamentals and
their overtones.  
The spectral region of the torsional overtones is further complicated
due to their coupling with the COC bending mode
\citep{senent1995b},  and the presence of many hot bands.
The following notation will be used throughout the paper:
$\nu_{11}$ is the ``anti-geared'' torsion, 
$\nu_{15}$ the ``geared'' torsion, 
and $\nu_7$ the COC bending mode;
vib-torsional energy levels will be denoted as $(v_{11} \, v_{15} \, v_7)$,
where $v_n$ is the number of torsional or vibrational quanta in the corresponding mode.

The torsional spectrum of DME was explored experimentally at room temperature
by~\citet{groner1977},  using Raman and infrared techniques. 
The torsional fundamental $\nu_{15}$  was observed at 241.0 cm$^{-1}$. 
The other torsional fundamental $\nu_{11}$ is forbidden in the absorption spectrum, and
was estimated between 199 cm$^{-1}$ and 202 cm$^{-1}$.
The two torsional overtones were assigned to the peaks observed at
$2\nu_{11} = 395.5$~cm$^{-1}$ and $2\nu_{15}= 481.2$~cm$^{-1}$.
The lack of an accurate value for the $\nu_{11}$ torsional mode, prevented the 
determination of some of the interaction parameters of the
effective Hamiltonian, needed for further analysis. 

In this paper, the torsional Raman spectra of DME and of $^{13}$C-DME have been
recorded at room temperature, and of cooled DME in supersonic
jet. The spectrum of cooled DME allowed us to assign unequivocally
the torsional overtones and their first hot bands, amending some of the
previous assignments. In turn, the new frequencies have been
used to refine 3D quantum calculations employing 
state-of-the art CCSD(T) ab initio calculations~\citep{villa2011} and
a torsion-torsion-bending Hamiltonian~\citep{senent1995b,carvajal2012}. 
The new characterization provides improved values of the Hamiltonian parameters for
the analysis of DME and $^{13}$C-DME rotational spectra and
future astrophysical detections.

\section{New experimental Raman measurements}
\label{sec-Experiment}

Two sets of Raman spectra of DME were recorded in this work: i) at
room temperature, and ii) jet-cooled. The sample of DME was
supplied by PRAXAIR, nominal purity 99.8~\%, while that of $^{13}$C-DME 
was synthesized in the University of Kassel, 
  Germany~\citep{kutzer2016}; the latter sample was distilled by
liquid nitrogen to remove residual H$_2$ from the synthesis. Raman
spectra of DME and of $^{13}$C-DME were recorded at room temperature under static conditions (pressure
550~mbar and 380~mbar, respectively). In addition, supersonic jets of
mixtures of DME diluted in He were produced, to cool the DME. Raman
shifts were calibrated against Hg and Ne emission lines, and are
accurate to $\pm 0.1$~cm$^{-1}$. Raman scattering was excited by 10~W
of linearly polarized radiation at 532 nm from a Coherent Verdi V10
laser, sharply focused down to a 15~$\mu$m beam waist on the sample by
a $f$=35~mm lens. Scattered radiation perpendicular to both laser
propagation and polarization was collected by an $f$=55 mm
photographic objective (Nikon, $f$/1.8) and projected, with a total
magnification $\times$10, onto the entrance slit of the
spectrograph. This is a Jobin-Yvon double monochromator, equipped with
two 2400 groove/mm gratings, and a CCD detector with
13$\times$13~$\mu \rm{m}^2$ pixels, refrigerated by liquid nitrogen. Entrance
slit was 75 and 150~$\mu$m, yielding spectral resolution of 0.36 and
0.72~cm$^{-1}$, respectively. Several scans were spike-filtered and averaged. 

\begin{figure}
  \includegraphics*[width=1.0\textwidth]{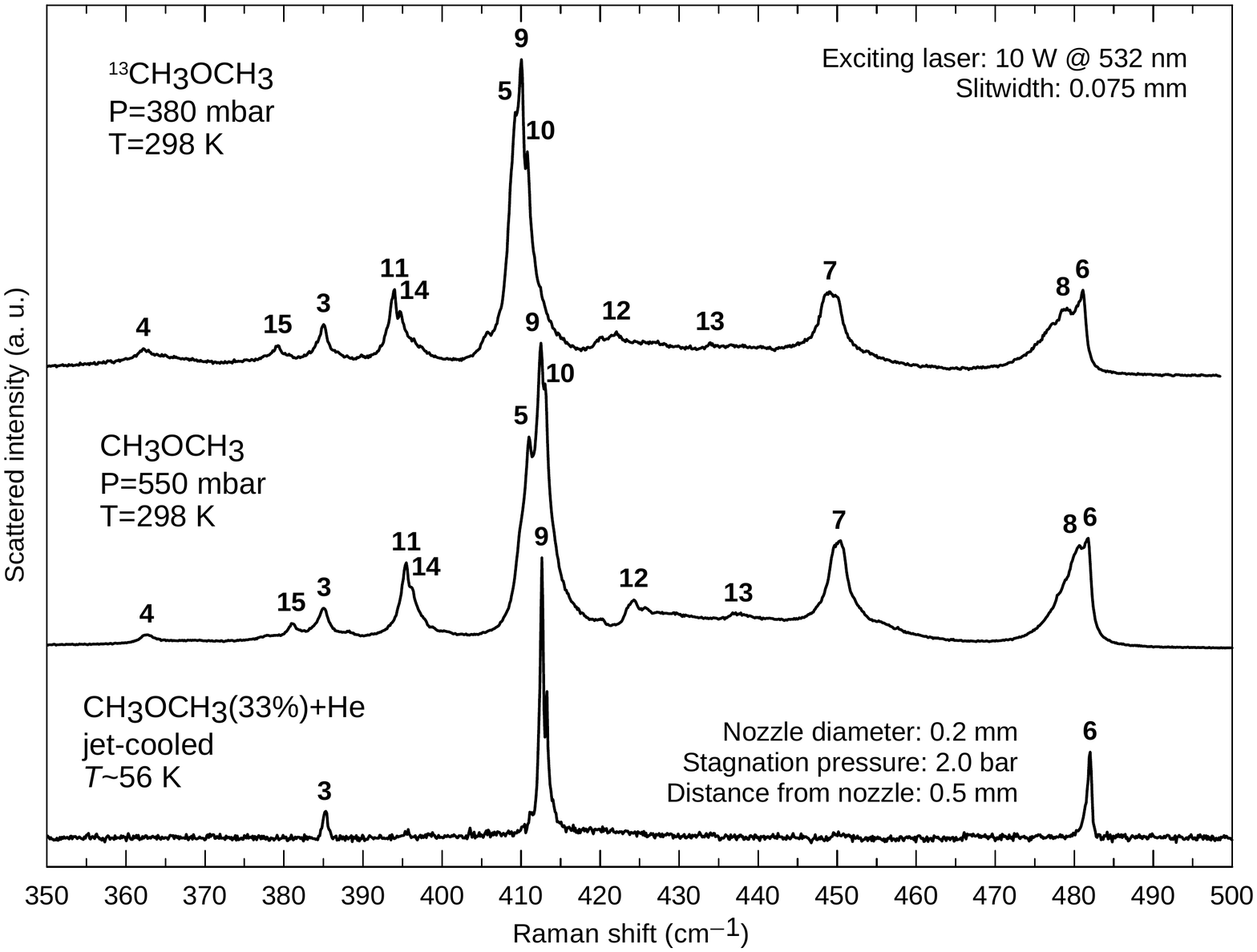}  
\caption{Raman spectrum of CH$_3$OCH$_3$ and $^{13}$CH$_3$OCH$_3$ in the region of the torsional overtones.
Band numbering refer to Table~\ref{tab_RamanTrans} (Appendix).}
 \label{fig1}
\end{figure}

\begin{figure}
  \includegraphics*[width=1.0\textwidth]{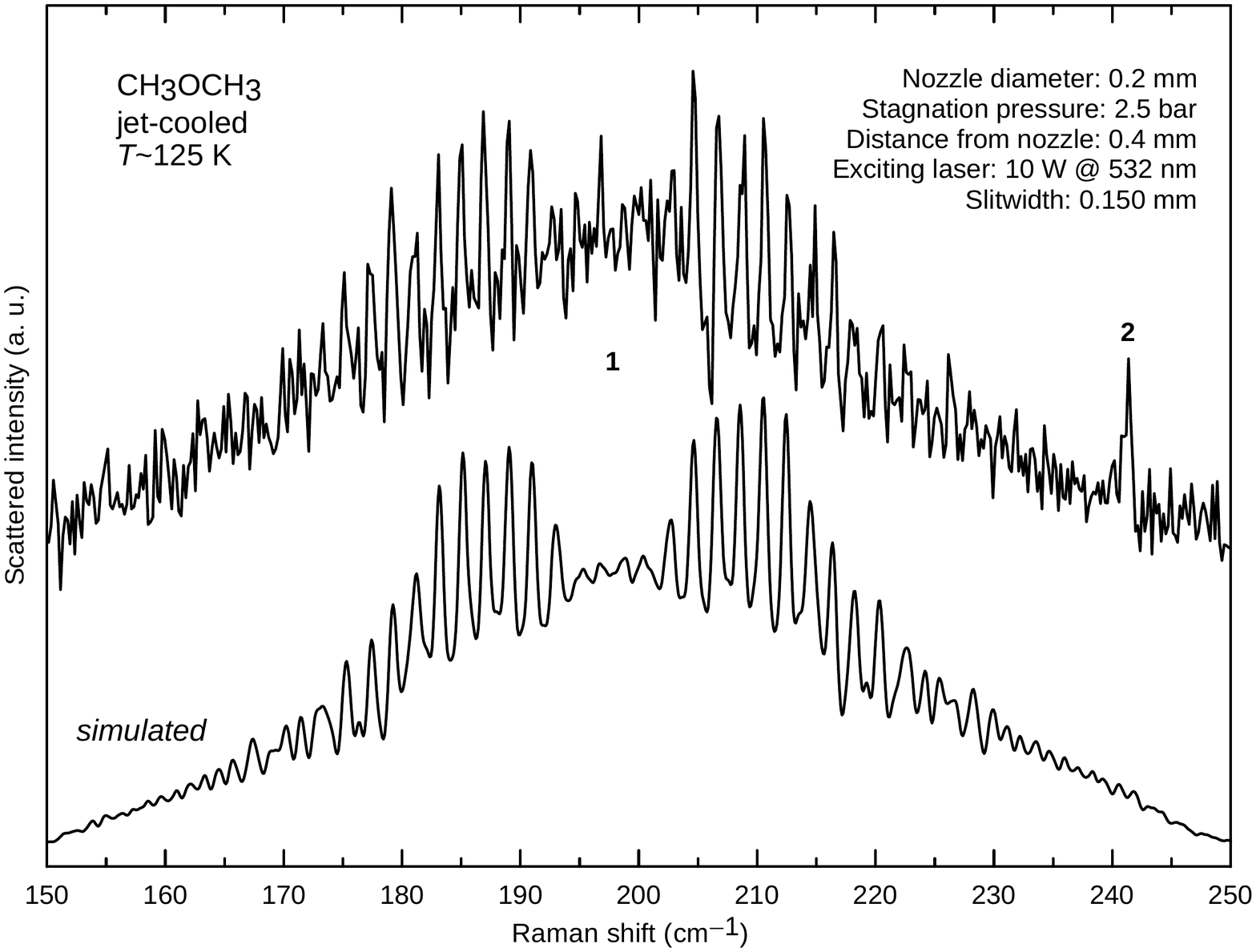}  
\caption{Raman spectrum of jet-cooled CH$_3$OCH$_3$ in the region of the torsional fundamentals.
Band numbering refer to Table~\ref{tab_RamanTrans} (Appendix).}
 \label{fig2}
\end{figure}

Raman spectroscopy can be an invaluable tool to investigate the
torsional modes of symmetric molecules like ethane \citep{fant86,jmf89b}, 
propane \citep{enge90},  or butane \citep{comp80,enge91}.
In such symmetric molecules, some of the
torsional modes are forbidden in IR or MW absorption, while the
torsional overtones give rise to weak Q-branches in Raman spectrum,
often accompanied by a rich structure of hot bands. 
In the case of DME, the $\nu_{11}$ torsional mode is silent in IR-MW; on the contrary,
the two torsional modes are Raman allowed, as well as their
overtones. 
The Raman spectra of DME and $^{13}$C-DME in this latter region is shown in
Fig.~\ref{fig1}. This region is dominated by the COC bending mode
$\nu_7$, which yields the peak at $\sim 410$~cm$^{-1}$.
In addition to that, a rich structure of peaks due to hot bands can be
seen, what difficults the safe assignment of the two torsional
overtones. The spectrum of $^{13}$C-DME is rather similar to that of DME,
with some of the peaks shifted towards lower wavenumbers, due to the
mass increase, as discussed below. The lowest panel shows the spectrum
of jet cooled DME in a supersonic expansion diluted in helium. It can
be clearly seen that at low temperature, only three peaks survive,
which can be assigned unambiguously to the two torsional overtones
plus the COC bending modes. Actually, the $2\nu_{11}$ torsional
overtone had been wrongly assigned \citep{groner1977} to the peak at
395.5~cm$^{-1}$ in the room temperature spectrum, which dissapears when
the molecule is jet-cooled, while the peak at 385.2~cm$^{-1}$ remains
at low temperature. 

The Raman spectrum of the jet-cooled DME in the region of the
torsional fundamentals is shown in Fig.~\ref{fig2}. 
These Raman bands are extremely weak, and had not been reported before. 
The faint Q-branch at 241.4~cm$^{-1}$ is the $\nu_{15}$
mode, which was observed previously in IR~\citep{fateley1962,groner1977}.
The broad band with the comb-like rotational structure can be assigned
to the unobserved $\nu_{11}$ torsional fundamental. 
A simulation with PGOPHER~\citep{western2016}, with the selection rules $\Delta K_a=\pm 1$
and $\Delta K_c=\pm 1$, corresponding to a  
Raman band of A$_2$ symmetry in the $C_{2v}$ point group, allows to
locate the band origin at 197.8~cm$^{-1}$,
although it could be downshifted by one unit of the peak spacing (1.9~cm$^{-1}$).

\section{Theoretical model for the spectral analysis}
\label{sec-theor}

In previous papers on DME~\citep{villa2011} and its isotopologues
 CH$_3$OCD$_3$ and CD$_3$OCD$_3$~\citep{senent2012},
$^{13}$CH$_3$OCH$_3$~\citep{carvajal2012}, and
CH$_3$OCH$_2$D~\citep{carvajal2014},
we used a three-dimensional (3D) model for the
analysis of their far infrared spectra. 
Assuming that the three low-frequency vibrational modes can be treated  separately
from the remaining ``high-frequency'' vibrations,
the 3D-Hamiltonian can be written as:   
\begin{equation}
\label{hamilt}
{\hat H}(\alpha,\theta_1,\theta_2) = - \sum_{i=1}^3 \, \sum_{j=1}^3 \,
\frac{\partial}{\partial q_i} B_{i j}(\alpha,\theta_1,\theta_2)
\frac{\partial}{\partial q_j} 
+ V^{\rm eff}(\alpha,\theta_1,\theta_2)
\end{equation}
In this equation, $q_i, q_j = (\alpha,\theta_1,\theta_2)$ represent
the three independent variables: the COC bending $\alpha$ and the two torsional
coordinates $\theta_1, \theta_2$.  The first term, which depends on the $B_{i j}$
parameters (the G matrix elements in cm$^{-1}$), is a 3D-kinetic energy operator.
The second term, the effective potential energy
$V^{\rm eff}$, is the sum of three contributions
\begin{equation}
\label{PES}
V^{\rm eff}(\alpha,\theta_1,\theta_2)=V(\alpha,\theta_1,\theta_2)
 + V^{\prime}(\alpha,\theta_1,\theta_2)+  V^{\rm ZPVE}(\alpha,\theta_1,\theta_2)
\end{equation}

\noindent where $V(\alpha,\theta_1,\theta_2)$
is the ab initio potential energy, 
$V^{\prime}(\alpha,\theta_1,\theta_2)$ the Podolsky pseudopotential,
and $V^{\rm ZPVE}(\alpha,\theta_1,\theta_2)$  
the zero point vibrational energy correction. 
The ab initio potential energy $V$ is isotopically invariant,
whereas $V^{\prime}$ and $V^{\rm ZPVE}$ 
(and thus the effective potential $V^{\rm eff}$)  depend on the nuclear masses,
as well as, of course, the kinetic energy parameters $B_{i j}$.

All the terms in Eqs.~(\ref{hamilt}) and (\ref{PES}) can be determined from energies, geometries, and harmonic frequencies from accurate ab initio calculations, as described elsewhere \citep{senent1998a, senent1998b}.
In the present paper, we used the previous ab initio calculations by~\citet{villa2011},
performed with the Gaussian 09 code~\citep{Gaussian}.
Coupled-cluster theory with single and double substitutions (CCSD)~\citep{scuseria1989}
was employed for the geometry optimizations. 
To improve the energies, single point
calculations were performed adding a perturbative treatment of triple excitations
(CCSD(T))~\citep{pople1987} on the CCSD geometries. Long range effects
are well described because the augmented aug-cc-pVTZ basis set was employed in all the
computations~\citep{woon1993}. 

Ab initio energies $V(\alpha,\theta_1,\theta_2)$ were calculated at 126 configurations for selected values of the three $(\alpha,\theta_1,\theta_2)$ coordinates: $\alpha = (104.676^{\circ} \rightarrow 119.676^{\circ}, \, \Delta \alpha= 3^{\circ})$, and $\theta_1,\theta_2=0, \pm30, \pm90, \pm150, 180$ degrees. 
In all the 126 configurations, the remaining $3N-9$ internal coordinates were allowed to relax. 
The other terms of the Hamiltonian, $B_{i j}$, $V'$, and $V^{\rm ZPVE}$, were also computed for the same 126 configurations; the $V^{\rm ZPVE}$ correction, needed to obtain reliable results~\citep{Smeyers1996}, was computed within the harmonic approximation.
Analytical  3D effective potential energy surfaces $V^{\rm eff}(\alpha,\theta_1,\theta_2)$ for DME and $^{13}$C-DME were then obtained by fitting their values at the 126 configurations to a series of the form
\begin{equation}
\label{Veff}
V^{\rm eff}(\alpha,\theta_1,\theta_2) = \sum_{l=0}^3 \left[ \sum_{m=0}^2 \sum_{n=0}^2
 A_{lmn}  \, \alpha^l \, \cos (3 m \theta_1) \cos (3 n \theta_2) + B_{l11}
\, \alpha^l \, \sin (3 \theta_1) \sin (3 \theta_2) \right]
\end{equation} 
For DME, $A_{lmn}= A_{lnm}$ and $B_{lmn}= B_{lnm}$,
while for the less symmetric $^{13}$C-DME, 
the $|A_{lmn} - A_{lnm}|$, and $|B_{lmn}- B_{lnm}|$ differences are found
to be lower than $0.0001$~cm$^{-1}$.
Formally identical fits were carried out for each of the kinetic energy operators $B_{i j}(\alpha,\theta_1,\theta_2)$.

The original code ENEDIM  \citep{senent2001} was used to carry out the calculations and fits of the 
 $V^{\rm eff}(\alpha,\theta_1,\theta_2)$ and $B_{i j}(\alpha,\theta_1,\theta_2)$ surfaces, 
and then to compute variationally the torsional and bending energy levels of the 3D Hamiltonian.
To reduce the computational expenses and for the classification of the levels, the molecular symmetry properties were taken into consideration. 
The two isotopologues studied here, $^{12}$CH$_3$O$^{12}$CH$_3$ (DME) and 
$^{13}$CH$_3$O$^{12}$CH$_3$ ($^{13}$C-DME), can be classified in the
$G_{36}$ and $G_{18}$ Molecular Symmetry Groups \citep{Bunker89}, respectively. 
$G_{36}$ contains nine irreducible representations: 
four non-degenerate $A_1$, $A_2$, $A_3$, and $A_4$, four doubly-degenerate $E_1$, $E_2$, $E_3$, 
and $E_4$ and one four-fold degenerate G. 
$G_{18}$ contains two non-degenerate representations $A_1$ and $A_2$ and four
double-degenerate ones $E_1$, $E_2$, $E_3$, and $E_4$. The correlation
between $G_{36}$ and $G_{18}$ representations was detailed in
previous papers~\citep{carvajal2012,Villa2013}. 

In our previous papers ~\citep{villa2011,carvajal2014}, the parameters of the ab initio
Hamiltonian were refined to reproduce the old
available experimental frequencies  of \citet{groner1977}.
In the present work, the refinement
has been revisited using the new Raman observations, as described
in the next section. We can anticipate that the ab initio
calculations are closer to the new experimental data than to the old ones.

\section{Discussion}

The new Raman spectra of cooled DME 
reported here (see Figs.~\ref{fig1} and~\ref{fig2})  probed crucial for  
the right assignment of the torsional modes and their overtones.
At $T=56$~K, only three bands survive in the recorded Raman spectrum
of DME in the bottom trace of Fig.~\ref{fig1}.
At such low $T$, only transitions starting from the ground state (000)
can be observed, due to the population distribution.
Thus, the observed 3 peaks can be safely assigned to
the transitions from ground state to
the (200), (001) and (020) levels, respectively. 
This has allowed us to reassign the torsional overtone (200) and,
with the help of the ab initio calculations, 
to amend the assignment of other bands 
already reported in both infrared and Raman spectra. 
In fact, \citet{groner1977} proposed that the torsional
overtone (200) of DME lie at  395.5~cm$^{-1}$
and its first hot band (100)$\rightarrow$(300) at 385.0~cm$^{-1}$
whereas, in this work, they are set at 385.2~cm$^{-1}$ and 362.6~cm$^{-1}$,
respectively. 
These new assignments are downshifted by around $\sim10$~cm$^{-1}$ 
and $\sim22$~cm$^{-1}$ with respect to those by \citet{groner1977}.
The peak around 395~cm$^{-1}$ is assigned here to two overlapping 
hot bands of the $\nu_7$ bending: (010)$\rightarrow$(011) and (001)$\rightarrow$(002).
The remaining prominent feature at $\sim$450~cm$^{-1}$ in Fig.~\ref{fig1},
unassigned in \citet{groner1977}, is the hot band (100)$\rightarrow$(120).

The complete lists of the experimental wavenumbers for both DME and $^{13}$C-DME,
along with their assignments and calculated torsional splittings, are displayed in Table~\ref{tab_RamanTrans} (Appendix).
The accuracy of the reported wavenumbers is mainly determined by the line shape,
as a result of the (unresolved) rotational structure and/or torsional splittings.
Thus, the quoted uncertainties range from $\pm 0.2$~cm$^{-1}$ for isolated narrow bands
up to $\pm 0.5$~cm$^{-1}$ for broad, overlapped, or weak bands.
As a representative check, the (001) bending energy levels, 
observed here at 412.5~cm$^{-1}$ for DME and 410.0~cm$^{-1}$ for $^{13}$C-DME,
are consistent within experimental error with the more accurate values
(412.350~cm$^{-1}$ and 409.993~cm$^{-1}$) reported recently by \citet{kutzer2016}.

It is worth to discuss the isotopic shifts, from DME to $^{13}$C-DME, of
the peaks in Fig.~\ref{fig1} in light of this new assigment. The
larger isotopic shifts ($\sim 2.5$~cm$^{-1}$) are those of the COC bending mode (and their
overtones and combinations), which are directly affected by the
increased mass of one of the C atoms. On the contrary, in zeroth-order
approximation, the torsional modes, (due to the internal rotation of
the two methyl groups) should not be affected by the mass of the C
atoms, because they lie on the rotation axis, and thus do not
contribute to the rotor moment of inertia. Of course, coupling between
the bending and the torsional modes leads to the observed isotopic
shifts of the torsional overtones.
The $\nu_{15}$ mode is  
more strongly coupled to the bending of
the COC skeletal frame (as revealed by its higher frequency and
intensity) than the $\nu_{11}$ mode. 
Thus, the new assigment proposed here is further supported by the tiny
isotopic shift (0.2 -- 0.4~cm$^{-1}$) of the peaks involving
$\nu_{11}$, as opposed to those ($\sim$~1~cm$^{-1}$) involving $\nu_{15}$.

To obtain the best fit parameters of the effective Hamiltonian, we started from
the ab initio  potential energy surfaces and the ab initio kinetic
parameters calculated in our previous works
on DME \citep{villa2011} and $^{13}$C-DME \citep{carvajal2012}.
Those ab initio calculations provided a reasonable description of
the torsional features but failed to reproduce the bending fundamental $\nu_7$ 
(421.64~cm$^{-1}$ ab initio vs 412.35~cm$^{-1}$ observed by \citet{kutzer2016}).
The ab initio Hamiltonian was then refined in this work following
three steps: i) the adjustment of the bending fundamental close to the experimental value;
ii) the reassignment of the two torsional fundamentals and their
 overtones according to the new Raman spectra;
iii) the assignment of the other observed bands and the subsequent global fit of the Hamiltonian.
The main difference, between the old Hamiltonians from~\citet{villa2011} and
\citet{carvajal2012} and the present one, involves the second step.
These three steps are detailed next.

i) The main weakness of the ab initio potential energy surface \citep{villa2011}
concerns the employed definition of the COC-bending coordinate,
which was set to the
COC angle $\alpha$, missing the contribution of other coordinates such
as the in-plane HCO angles. 
Therefore, for a more realistic description of the COC bending,
a new $\alpha^{\prime}$ coordinate was introduced
\begin{equation}
\label{alphap}
\alpha^{\prime} = \alpha \, (1 + F/100) ~~,
\end{equation}
where $F$ is a factor which corrects the contribution of the
curvilinear internal 
coordinate angle to the normal COC-bending coordinate.
Hence, the $B_{\alpha, \alpha}$ kinetic parameter
was also  corrected in all the conformations. For DME and $^{13}$C-DME, $F$ was optimized
to be 1.954~\citep{villa2011,carvajal2012} to 
reproduce the experimental bending energy term values. This
represents a coordinate correction lower than 2$\%$.

ii) 
The ab initio torsional overtones of DME~\citep{villa2011}, when compared 
with the experimental data by \citet{groner1977}, 
yielded the differences (obs-calc)
$\Delta=395.5 - 388.61 =+6.89$~cm$^{-1}$ for $2\nu_{11}$, 
and
$\Delta=481.2 -487.22= -6.02$~cm$^{-1}$ for $2\nu_{15}$.
One torsional overtone seemed to be underestimated 
by the calculations whereas the other one appeared overestimated. 
In the subsequent adjustment~\citep{villa2011},
the potential term $B_{011}$ in Eq.~\ref{Veff}, one of the main
responsible for the gap between $\nu_{11}$ and $\nu_{15}$ bands, 
had to be forced and strongly modified to reproduce the old data.

Here, with the new Raman reassignments, the 
discrepancies between experimental and ab initio values decrease significantly: 
$\Delta=385.2 -  388.61 =-3.41$~cm$^{-1}$ for $2\nu_{11}$, and 
$\Delta=482.0 - 487.22=-5.22$~cm$^{-1}$ for 2$\nu_{15}$.
Thus, both overtones appear now at higher frequencies in the calculations, 
and gaps between torsional levels are described correctly
without any refinement of the $B_{011}$ parameter.

iii) To reproduce the newly assigned experimental bands,
just a few parameters of the Hamiltonian need to be optimized:
one kinetic parameter and two potential parameters.
In Table~\ref{tab_PES_param} in the Appendix the fitted effective potential coefficients
($A_{200}$,  $A_{020}$) are indicated.

The new optimized Hamiltonians are more confident to be used in
the assignments of other bands of the two
isotopologues and in future works. For example, the infrared-forbidden torsional fundamental
$\nu_{11}$ of DME, calculated ab initio at 199.16~cm$^{-1}$ is predicted now at 198.33~cm$^{-1}$,
much closer to the present experimental observation at 197.8~cm$^{-1}$ (see Figure~\ref{fig2}). 

Table~\ref{Etors} lists the torsional-bending energies for DME and $^{13}$C-DME
up to $\sim$860~cm$^{-1}$ calculated in this work,
and compared with those reported previously.
It should be stressed that the calculated energy levels could be labelled within 
a  $(v_{11} \, v_{15} \, v_7)$ scheme up to $\sim 825$~cm$^{-1}$, as listed in Table~\ref{Etors}.
For higher energies, the large torsional splittings and mixing of the wavefunctions
impedes to assign unambiguously such a $(v_{11} \, v_{15} \, v_7)$ label to all the computed levels,
especially for the less symmetric $^{13}$C-DME.
The experimental Raman transition wavenumbers of DME and $^{13}$C-DME,
 and their assignments, are listed in Table~\ref{tab_RamanTrans} (Appendix) 
to facilitate the understanding of Figures~\ref{fig1} and \ref{fig2},
along with the calculated wavenumbers from the fitted energies of Table~\ref{Etors}.

\section{Concluding remarks}

The spectrum of DME and isotopologues in the THz region 
is rather complex due to the high density of states, the torsional splittings,
the coupling of the torsional overtones with the COC bending mode,
and the presence of many hot bands.
Thus, a conclusive assignment of the far infrared spectrum 
is greatly facilitated by measurements at different temperatures to distinguish
between cold and hot bands. This is one of the main results of the present work.
    
New laboratory measurements of the torsional Raman spectrum of DME and $^{13}$C-DME,
 from room temperature down to  56~K, are reported.
This has allowed us to observe the torsional band $\nu_{11}= 197.8$~cm$^{-1}$, not reported to date,
 and to reassign its overtone 
at $2\nu_{11}= 385.2$~cm$^{-1}$ and 385.0~cm$^{-1}$ for DME and $^{13}$C-DME, respectively. 
In due turn, this has also allowed us to assign correctly all the lowest energy levels (see
Table~\ref{Etors}), which are those relevant for the interpretation of
the astronomical observations. 

The new Raman measurements have been interpreted with the
help of highly correlated ab initio calculations within a 3D torsional-vibrational model.
The ab initio parameters of the 3D Hamiltonian
have been refined using the new experimental data.
In the past, such refinement standed out some problems derived from the lack
of experimental data corresponding to the $\nu_{11}$ torsional mode,
and the large density of states in the
region of the torsional overtones, where the coupling with the COC bending occur.  
These problems have been fixed here, reaching a better agreement with the experiment.
Eventually, the quantitative interpretation of the energy level structure in molecules with such large
amplitude internal motions  relies on
quantum chemical calculations validated by laboratory data.  

DME has been observed in excited torsional states in hot astrophysical environments.  
The present results can help to the characterization of low energy states 
which can be responsible for unidentified lines of astrophysical surveys,
where the presence of the low energy overtones can be relevant.
The best fit Hamiltonian reported here can be used to verify former spectral
analyses in the gas phase, and to predict other yet unobserved bands,
for future astrophysical detections of other isotopologues of DME in the ISM.

\acknowledgments

This research is supported by the Spanish Ministerio de Econom\'{i}a y
Competitividad (MINECO) under Grants FIS2014-53448-C2-2-P, FIS2016-76418-P and FIS2017-84391-C2. The authors also acknowledge the COST Actions CM1401 ``Our Astrochemical History'' and CM1405 ``MOLIM'' and CTI (CSIC) and CESGA for computing facilities. 
Thanks are due to Pia Kutzer and Thomas Giesen (Universit\"at Kassel) for providing the sample of $^{13}$CH$_3$OCH$_3$ used in this work.

%%%%%%%%%%%%%%%%%%%%%%%%%%%%%%%%%%%%
\begin{deluxetable}{ccrrcrr}
\tabletypesize{\scriptsize}
\tablecaption{Energies (in cm$^{-1}$) of the lowest torsional and bending levels
of DME and $^{13}$C-DME.
\label{Etors}}
\tablewidth{0pt}
\tablehead{
$(v_{11} \, v_{15} \, v_7)$ & Symm. &\multicolumn{2}{c}{$^{12}$CH$_3$O$^{12}$CH$_3$} & Symm. & \multicolumn{2}{c}{$^{13}$CH$_3$O$^{12}$CH$_3$}\\
\hline
\colhead{}&
\colhead{} & \colhead{OLD\tablenotemark{a}} & \colhead{NEW\tablenotemark{b}} &
\colhead{} & \colhead{OLD\tablenotemark{c}} & \colhead{NEW\tablenotemark{b}}
}
\startdata 
$0 \, 0 \, 0$ & A1 &   0.000 &     0.000       &  A1 &   0.000 &       0.000 \\
              & G  &   0.000 &     0.000       &  E1 &   0.000 &       0.000 \\
              & E1 &   0.001 &     0.001       &  E2 &   0.000 &       0.000 \\
              & E3 &   0.001 &     0.001       &  E3 &   0.001 &       0.001 \\
              &    &         &                 &  E4 &   0.001 &       0.001 \\
 \hline                                                        
$1 \, 0 \, 0$ & A3 & 201.611 &    198.341      &  A2 & 200.912 &     197.359 \\
              & G  & 201.602 &    198.332      &  E1 & 200.903 &     197.350 \\
              & E2 & 201.593 &    198.323      &  E2 & 200.904 &     197.350 \\
              & E3 & 201.593 &    198.323      &  E3 & 200.895 &     197.341 \\
              &    &         &                 &  E4 & 200.895 &     197.341 \\
 \hline                                                        
$0 \, 1 \, 0$ & A2 & 241.783 &    242.603      &  A2 & 241.607 &     242.686 \\
              & G  & 241.774 &    242.593      &  E1 & 241.598 &     242.676 \\
              & E1 & 241.765 &    242.584      &  E2 & 241.598 &     242.676 \\
              & E4 & 241.765 &    242.584      &  E3 & 241.589 &     242.667 \\
              &    &         &                 &  E4 & 241.589 &     242.667 \\
 \hline                                                        
$2 \, 0 \, 0$ & A1 & 391.094 &    386.528      &  A1 & 389.609 &     384.831 \\
              & G  & 391.225 &    386.658      &  E1 & 389.732 &     384.954 \\
              & E1 & 391.358 &    386.789      &  E2 & 389.734 &     384.956 \\
              & E3 & 391.358 &    386.789      &  E3 & 389.859 &     385.081 \\
              &    &         &                 &  E4 & 389.859 &     385.081 \\
 \hline                                                        
$0 \, 0 \, 1$ & A1 & 412.086 &    413.042      &  A1 & 409.170 &     410.432 \\
              & G  & 412.086 &    413.039      &  E1 & 409.173 &     410.430 \\
              & E1 & 412.087 &    413.036      &  E2 & 409.173 &     410.430 \\
              & E3 & 412.087 &    413.036      &  E3 & 409.176 &     410.428 \\
              &    &         &                 &  E4 & 409.176 &     410.428 \\
 \hline                                                        
$1 \, 1 \, 0$ & A4 & 422.176 &    420.471      &  A1 & 421.629 &     419.905 \\
              & G  & 422.393 &    420.691      &  E1 & 421.846 &     420.123 \\
              & E2 & 422.609 &    420.910      &  E2 & 421.838 &     420.116 \\
              & E4 & 422.609 &    420.910      &  E3 & 422.053 &     420.333 \\
              &    &         &                 &  E4 & 422.054 &     420.333 \\
 \hline                                                        
$0 \, 2 \, 0$ & A1 & 480.889 &    481.940      &  A1 & 479.987 &     481.654 \\
              & G  & 480.960 &    482.020      &  E1 & 480.059 &     481.734 \\
              & E1 & 481.031 &    482.099      &  E2 & 480.057 &     481.731 \\
              & E3 & 481.031 &    482.099      &  E3 & 480.129 &     481.811 \\
              &    &         &                 &  E4 & 480.129 &     481.811 \\
 \hline                                                        
$3 \, 0 \, 0$ & A3 & 569.642 &    564.130      &  A2 & 568.029 &     562.043 \\
              & G  & 567.967 &    562.616      &  E1 & 566.480 &     560.650 \\
              & E2 & 566.614 &    561.352      &  E2 & 566.406 &     560.592 \\
              & E3 & 566.615 &    561.352      &  E3 & 565.143 &     559.416 \\
              &    &         &                 &  E4 & 565.143 &     559.416 \\
 \hline                                                        
$2 \, 1 \, 0$ & A2 & 590.656 &    587.974      &  A2 & 589.896 &     587.047 \\
              & G  & 588.364 &    585.674      &  E1 & 587.584 &     584.752 \\
              & E1 & 585.795 &    583.182      &  E2 & 587.703 &     584.850 \\
              & E4 & 585.797 &    583.184      &  E3 & 585.152 &     582.397 \\
              &    &         &                 &  E4 & 585.153 &     582.398 \\
 \hline                                                        
$1 \, 0 \, 1$ & A3 & 613.117 &    611.569      &  A2 & 610.202 &     608.235 \\
              & G  & 613.206 &    611.610      &  E1 & 610.290 &     608.283 \\
              & E2 & 613.294 &    611.643      &  E2 & 610.291 &     608.286 \\
              & E3 & 613.294 &    611.642      &  E3 & 610.379 &     608.329 \\
              &    &         &                 &  E4 & 610.378 &     608.329 \\
 \hline                                                        
$0 \, 1 \, 1$ & A2 & 636.395 &    638.255      &  A2 & 634.044 &     636.816 \\
              & G  & 636.449 &    638.313      &  E1 & 634.095 &     636.870 \\
              & E1 & 636.506 &    638.371      &  E2 & 634.090 &     636.867 \\
              & E4 & 636.507 &    638.372      &  E3 & 634.144 &     636.922 \\
              &    &         &                 &  E4 & 634.144 &     636.923 \\
 \hline                                                        
$1 \, 2 \, 0$ & A3 & 648.652 &    647.438      &  A2 & 646.719 &     645.721 \\
              & G  & 647.493 &    646.185      &  E1 & 645.527 &     644.432 \\
              & E2 & 646.295 &    644.894      &  E2 & 645.564 &     644.471 \\
              & E3 & 646.286 &    644.884      &  E3 & 644.328 &     643.140 \\
              &    &         &                 &  E4 & 644.319 &     643.129 \\
 \hline                                                        
$4 \, 0 \, 0$ & A1 & 706.250 &    703.223      &  A1 & 705.812 &     702.388 \\
              & G  & 708.488 &    705.787      &  E1 & 707.928 &     704.883 \\
              & E1 & 727.268 &    723.674      &  E2 & 708.391 &     705.332 \\
              & E3 & 726.577 &    722.224      &  E3 & 726.116 &     722.212 \\
              &    &         &                 &  E4 & 725.509 &     720.574 \\
 \hline                                                        
$0 \, 3 \, 0$ & A2 & 717.917 &    718.784      &  A2 & 716.038 &     718.141 \\
              & G  & 717.442 &    718.122      &  E1 & 715.594 &     717.483 \\
              & E1 & 716.716 &    716.717      &  E2 & 715.601 &     717.481 \\
              & E4 & 717.059 &    717.671      &  E3 & 714.923 &     715.883 \\
              &    &         &                 &  E4 & 715.225 &     717.040 \\
 \hline                                                        
$3 \, 1 \, 0$ & A4 & 711.733 &    709.884      &  A1 & 711.676 &     709.692 \\
              & G  & 731.779 &    728.819      &  E1 & 731.382 &     728.093 \\
              & E2 & 735.657 &    733.397      &  E2 & 730.719 &     727.475 \\
              & E4 & 736.031 &    733.928      &  E3 & 735.199 &     732.699 \\
              &    &         &                 &  E4 & 735.531 &     733.219 \\
 \hline                                                        
$2 \, 2 \, 0$ & A1 & 785.149 &    779.346      &  A1 & 783.614 &     777.555 \\
              & G  & 784.508 &    779.640      &  E1 & 783.133 &     778.089 \\
              & E1 & 783.988 &    779.822      &  E2 & 783.084 &     778.032 \\
              & E3 & 783.912 &    779.724      &  E3 & 782.666 &     778.371 \\
              &    &         &                 &  E4 & 782.595 &     778.274 \\
 \hline                                                        
$0 \, 0 \, 2$ & A1 & 805.550 &    806.600      &  A1 & 800.434 &     802.112 \\
              & G  & 805.861 &    806.826      &  E1 & 800.978 &     802.594 \\
              & E1 & 806.194 &    807.097      &  E2 & 800.964 &     802.564 \\
              & E3 & 806.105 &    806.966      &  E3 & 801.516 &     803.083 \\
              &    &         &                 &  E4 & 801.443 &     802.962 \\
 \hline 
$1 \, 3 \, 0$ & A4 & 810.976 &    810.464      &  A1 & 809.108 &     808.658 \\
              & G  & 809.429 &    808.742      &  E1 & 807.652 &     807.050 \\
              & E2 & 807.515 &    806.719      &  E2 & 807.701 &     807.095 \\
              & E4 & 807.579 &    806.834      &  E3 & 805.916 &     805.227 \\
              &    &         &                 &  E4 & 805.970 &     805.333 \\
 \hline                                                        
$2 \, 0 \, 1$ & A1 & 823.691 &    824.164      &  A1 & 819.009 &     819.193 \\
              & G  & 822.370 &    822.840      &  E1 & 818.451 &     818.710 \\
              & E1 & 821.995 &    822.630      &  E2 & 818.364 &     818.633 \\
              & E3 & 821.689 &    822.259      &  E3 & 818.138 &     818.488 \\
              &    &         &                 &  E4 & 817.920 &     818.254 \\
 \hline Unlabelled energy levels & & \vdots & \vdots & & \vdots & \vdots \\
 \hline                                                        
$0 \, 2 \, 1$\tablenotemark{d} & A1 & 858.148 & 860.357 &  A1 & 855.085 & 858.393 \\
              & G  & 857.950 & 860.106 &  E1 & 854.896 & 858.156 \\
              & E1 & 857.832 & 859.899 &  E2 & 854.913 & 858.173 \\
              & E3 & 857.778 & 859.845 &  E3 & 854.795 & 857.961 \\
              &    &         &         &  E4 & 854.746 & 857.914 \\
\hline
\enddata

\tablenotetext{a}{Calculated with the adjusted Hamiltonian of \citet{villa2011}.}
\tablenotetext{b}{Calculated with the refined Hamiltonian of this work.}
\tablenotetext{c}{Calculated with the adjusted Hamiltonian of \citet{carvajal2012}.}
\tablenotetext{d}{Tentative label.}

\end{deluxetable}

%\appendix

%\section*{A. Appendix information}

%This Appendix contains two tables, one with the experimental and calculated wavenumbers for the Raman transitions of DME and $^{13}$C-DME and their assignments, 
%and another one with the coefficients for their refined effective Potential Energy Surface.

\begin{deluxetable}{cccccccc}
\tabletypesize{\scriptsize}
\tablecaption{APPENDIX: Experimental and calculated wavenumbers (cm$^{-1}$) of the
 Raman transitions of DME and $^{13}$C-DME.\label{tab_RamanTrans}}
\tablewidth{0pt}
\tablehead{ Band \# &
$(v_{11} \, v_{15} \, v_7)_i$ $\to$ $(v_{11} \, v_{15} \, v_7)_f$&Symm.&\multicolumn{2}{c}{$^{12}$CH$_3$O$^{12}$CH$_3$}&Symm.&\multicolumn{2}{c}{$^{13}$CH$_3$O$^{12}$CH$_3$}\\
\hline
\colhead{}& \colhead{}& \colhead{} & \colhead{Calc.\tablenotemark{a}} & \colhead{Exp.\tablenotemark{b}} &
\colhead{} &\colhead{Calc.\tablenotemark{a}} & \colhead{Exp.\tablenotemark{b}}
}
\startdata 
1 & $0 \, 0 \, 0 \to 1 \, 0 \, 0$ 
              & A3  &   198.34  & $197.8 \pm 0.2$  &  A2 & 197.36 & -\\
            & & G   &   198.33  &        &  E1 & 197.35 \\
            & & E2  &   198.32  &        &  E2 & 197.35 \\
            & & E3  &   198.32  &        &  E3 & 197.34 \\
            & &     &           &        &  E4 & 197.34 \\
 \hline                                               
2 & $0 \, 0 \, 0 \to 0 \, 1 \, 0$
              & A2  &   242.60  & $241.4 \pm 0.4$ &  A2 & 242.69 & $241.4 \pm 0.5$ \\
            & & G   &   242.59  &        &  E1 & 242.68 \\
            & & E1  &   242.58  &        &  E2 & 242.68 \\
            & & E4  &   242.58  &        &  E3 & 242.67 \\
            & &     &           &        &  E4 & 242.67 \\
 \hline     \hline                                                   
3 & $0 \, 0 \, 0 \to 2 \, 0 \, 0$
              & A1  &   386.53  & $385.2 \pm 0.2$ & A1 &  384.83 & $385.0 \pm 0.2$ \\
            & & G   &   386.66  &        & E1 &  384.95 &  \\
            & & E1  &   386.79  &        & E2 &  384.96 &  \\
            & & E3  &   386.79  &        & E3 &  385.08 &  \\
            & &     &           &        & E4 &  385.08 &  \\
\hline
4 & $1 \, 0 \, 0 \to 3 \, 0 \, 0$
              & A3  &   365.79  & $362.6 \pm 0.2$ & A2 &  364.68 & $362.3 \pm 0.2$ \\
            & & G   &   364.28  &        & E1 &  363.30 &  \\
            & & E2  &   363.03  &        & E2 &  363.24 &  \\
            & & E3  &   363.03  &        & E3 &  362.08 &  \\
            & &     &           &        & E4 &  362.08 &  \\
\hline
5 & $0 \, 0 \, 1 \to 2 \, 0 \, 1$ 
              & A1  &   411.12  & $411.1 \pm 0.2$ & A1 &  408.76 & $409.3 \pm 0.2$ \\
            & & G   &   409.80  &        & E1 &  408.28 &  \\
            & & E1  &   409.59  &        & E2 &  408.20 &  \\
            & & E3  &   409.22  &        & E3 &  408.06 &  \\
            & &     &           &        & E4 &  407.83 &  \\
\hline\hline
6 & $0 \, 0 \, 0 \to 0 \, 2 \, 0$
              & A1  &   481.94 & $482.0 \pm 0.2$ & A1 &  481.65 & $481.1 \pm 0.2$  \\
            & & G   &   482.02 &         & E1 &  481.73 &  \\
            & & E1  &   482.10 &         & E2 &  481.73 &  \\
            & & E3  &   482.10 &         & E3 &  481.81 &  \\
            & &     &          &         & E4 &  481.81 &  \\
\hline
7 & $1 \, 0 \, 0 \to 1 \, 2 \, 0$
              & A3  &   449.10 & $450.2 \pm 0.4$ & A2 &  448.36 & $449.2 \pm 0.5$ \\
            & & G   &   447.85 &         & E1 &  447.08 &  \\
            & & E2  &   446.57 &         & E2 &  447.12 &  \\
            & & E3  &   446.56 &         & E3 &  445.80 &  \\
            & &     &          &         & E4 &  445.79 &  \\
\hline
8 & $0 \, 1 \, 0 \to 0 \, 3 \, 0$
              & A2  &   476.18 & $480.7 \pm 0.2$ & A2 &  475.46 & $479.0 \pm 0.3$  \\
            & & G   &   475.53 &         & E1 &  474.81&  \\
            & & E1  &   474.13 &         & E2 &  474.81&  \\
            & & E4  &   475.09 &         & E3 &  473.22&  \\
            & &     &          &         & E4 &  474.37&  \\
\hline\hline
9 & $0 \, 0 \, 0 \to 0 \, 0 \, 1$
              & A1  &   413.04 & $412.5 \pm 0.2$ & A1 & 410.43 & $410.0 \pm 0.2$ \\
            & & G   &   413.04 &         & E1 &  410.43 &  \\
            & & E1  &   413.04 &         & E2 &  410.43 &  \\
            & & E3  &   413.04 &         & E3 &  410.43 &  \\
            & &     &          &         & E4 &  410.43 &  \\
\hline
10 & $1 \, 0 \, 0 \to 1 \, 0 \, 1$
              & A3  &   413.23 & $413.2 \pm 0.2$ & A2 & 410.88 & $410.8 \pm 0.2$ \\
            & & G   &   413.28 &         & E1 &  410.93 &  \\
            & & E2  &   413.32 &         & E2 &  410.94 &  \\
            & & E3  &   413.32 &         & E3 &  410.99 &  \\
            & &     &          &         & E4 &  410.99 &  \\
\hline
11 & $0 \, 1 \, 0 \to 0 \, 1 \, 1$
              & A2  &   395.65 & $395.5 \pm 0.2$ & A2 & 394.13 & $393.9 \pm 0.2$ \\
            & & G   &   395.72 &         & E1 &  394.19 &  \\
            & & E1  &   395.79 &         & E2 &  394.19 &  \\
            & & E4  &   395.79 &         & E3 &  394.26 &  \\
            & &     &          &         & E4 &  394.26 &  \\
\hline
12 & $2 \, 0 \, 0 \to 0 \, 0 \, 2$
              & A1  &   420.07 & $424.2 \pm 0.2$ & A1 & 417.28 & $422.0 \pm 0.3$ \\
            & & G   &   420.17 &         & E1 &  417.64 &  \\
            & & E1  &   420.31 &         & E2 &  417.61 &  \\
            & & E3  &   420.18 &         & E3 &  418.00 &  \\
            & &     &          &         & E4 &  417.88 &  \\
\hline
13 & $2 \, 0 \, 0 \to 2 \, 0 \, 1$
              & A1  &   437.64 & $437.5 \pm 0.5$ & A1 & 434.36 & $434.0 \pm 0.2$ \\
            & & G   &   436.18 &         & E1 &  433.76 &  \\
            & & E1  &   435.84 &         & E2 &  433.68 &  \\
            & & E3  &   435.47 &         & E3 &  433.41 &  \\
            & &     &          &         & E4 &  433.17 &  \\
\hline
14 & $0 \, 0 \, 1 \to 0 \, 0 \, 2$
              & A1  &   393.56 & $396.3 \pm 0.3$ & A1 & 391.68 & $394.7 \pm 0.2$ \\
            & & G   &   393.79 &         & E1 &  392.16 &  \\
            & & E1  &   394.06 &         & E2 &  392.13 &  \\
            & & E3  &   393.93 &         & E3 &  392.66 &  \\
            & &     &          &         & E4 &  392.53 &  \\
\hline
15 & $0 \, 2 \, 0 \to 0 \, 2 \, 1$\tablenotemark{c} 
              & A1  &   378.42 & $381.0 \pm 0.2$ & A1 & 376.74 & $379.2 \pm 0.2$ \\
            & & G   &   378.09 &         & E1 &  376.42 &  \\
            & & E1  &   377.80 &         & E2 &  376.44 &  \\
            & & E3  &   377.75 &         & E3 &  376.15 &  \\
            & &     &          &         & E4 &  376.10 &  \\
\hline
\enddata

\tablenotetext{a}{Calculated with the refined Hamiltonian of this work.}
\tablenotetext{b}{Experimental.}
\tablenotetext{c}{Tentative assignment.}
\end{deluxetable}

\begin{deluxetable}{ccc}
\tabletypesize{\scriptsize}
%\rotate
\tablecaption{APPENDIX: Expansion Coefficients for the refined
  CCSD(T)/aug-cc-pVTZ effective Potential Energy Surface
  of $^{12}$CH$_3$O$^{12}$CH$_3$ and $^{13}$CH$_3$O$^{12}$CH$_3$.\label{tab_PES_param}}
\tablewidth{0pt}
\tablehead{
&$^{12}$CH$_3$O$^{12}$CH$_3$&$^{13}$CH$_3$O$^{12}$CH$_3$\\
\hline
\colhead{ Coeff.\tablenotemark{a}} & \colhead{V$^{\rm  eff}$~\tablenotemark{b}} & 
\colhead{V$^{\rm eff}$~\tablenotemark{b}} 
}
\startdata
$A_{000}$ & 970.1546& 970.1576\\
$A_{100}$ & -54.6202&-54.7897\\
$A_{200}$ &  10.2064$^{\rm fitted}$& 10.2329$^{\rm fitted}$\\
$A_{300}$ &  -0.1424& -0.1402\\
$A_{010}$ & -495.2177& -497.1729\\
$A_{110}$ & 31.7310& 31.7310\\
$A_{210}$ & -1.0805& -1.0805\\
$A_{310}$ & 0.0134& 0.0134\\
$A_{011}$ & 25.4803& 26.7661\\
$A_{111}$ & -12.7102& -12.7102\\
$A_{211}$ & 0.5208& 0.5208\\
$A_{311}$ & -0.0056& -0.0056\\
$A_{020}$ & -1.8845$^{\rm fitted}$& -0.9984$^{\rm fitted}$\\
$A_{120}$ & 1.1183& 1.1183\\
$A_{220}$ & -0.1071& -0.1071\\
$A_{320}$ & 0.0038& 0.0038\\
$A_{021}$ & 0.7500& 0.8149\\
$A_{121}$ & -0.3289& -0.3289\\
$A_{221}$ & 0.0481& 0.0481\\
$A_{321}$ & -0.0026&-0.0026 \\
$A_{022}$ & 1.7800& 1.3639\\
$A_{122}$ & -0.0958& -0.0958\\
$A_{222}$ & -0.0293& -0.0293\\
$A_{322}$ & 0.0026& 0.0026\\
$B_{011}$ & -13.9554& -15.8995\\
$B_{111}$ & 14.3670& 14.3670\\
$B_{211}$ & -0.7375& -0.7375\\
$B_{311}$ & 0.0117& 0.0117\\
\enddata

\tablenotetext{a}{Expansion coefficients $A_{lmn}$ and
  $B_{lmn}$ (in units of cm$^{-1}$/degrees$^l$) of Eq.~(\ref{Veff}).}
\tablenotetext{b}{Refined effective potential parameters. Fitted parameters are indicated.}
\end{deluxetable}
%%%%%%%%%%%%%%%%%%%%%%%%%%%%%%%%%%%%

\begin{thebibliography}{}
\bibitem[Balucani et al.(2015)]{balucani2015} Balucani, N.,
  Ceccarelli, C., \& Taquet, V. 2015, MNRAS, 449, L16
\bibitem[Bisschop et al.(2013)]{bisschop2013} Bisschop, S.E., Schilke,
  P., Wyrowski, F., Belloche, A., Brinch, C., Endres, C.P., Gusten,
  R., Hafok, H., Heyminck, S., Jorgensen, J.K., M\"uller, H.S.P.,
  Menten, K.M., Rolffs, R., \& Schlemmer, S. 2013, A\&A, 552, 122
\bibitem[Blukis et al.(1963)]{blukis1963} Blukis, U., Kasai, P.H., \&
  Myers, R.J. 1963, JChPh, 38, 2753 
\bibitem[Brouillet et al.(2013)]{brouillet2013} Brouillet, N.,
  Despois, D., Baudry, A., Peng, T.-C., Favre, C., Wootten, A.,
  Remijan, A.J., Wilson, T.L., Combes, F., \& Wlodarczak, G. 2013,
  A\&A, 550, 46
\bibitem[Bunker \& Jensen (1989)]{Bunker89} Bunker, P.~R. \& Jensen,
  P. 1989, {\em Molecular Symmetry and Spectroscopy},
  NRC Research Press, Ottawa.
\bibitem[Carvajal et al.(2010)]{carvajal2010} Carvajal, M., Kleiner,
  I., \& Demaison, J. 2010, \apjs, 190, 315
\bibitem[Carvajal et al.(2012)]{carvajal2012} Carvajal, M.,
  \'Alvarez-Bajo, O., Senent, M.L., Dom\'{\i}nguez-G\'omez, R., \&
  Villa, M. 2012, JMoSp, 279, 3 
\bibitem[Carvajal et al.(2014)]{carvajal2014} Carvajal, M., Senent,
  M.L., Villa, M., \& Dom\'{\i}nguez-G\'omez, R. 2014,
  CPL, 592, 200 
\bibitem[Compton et al. (1980)]{comp80}
  Compton, D.A.C., Montero, S., \& Murphy, W.F 1980,
  JPhCh, 84, 3587
\bibitem[Coudert et al.(2002)]{coudert2002} Coudert, L.H.,
  \c{C}ar\c{c}abal, P., Chevalier, M., Broquier, M., Hepp, M., \&
  Herman, M. 2002, JMoSp, 212, 203 
\bibitem[Durig et al.(1976)]{durig1976} Durig, J.R., Li, Y.S., \&
  Groner, P. 1976, JMoSp, 62, 159 
\bibitem[Endres et al.(2009)]{endres2009} Endres, C.P., Drouin, B.J.,
  Pearson, J.C., M\"uller, H.S.P., Lewen, F., Schlemmer, S., \& Giesen
  T.F. 2009, A\&A, 504, 635 
\bibitem[Engeln et al. (1990)]{enge90} 
  Engeln, R., Reuss, J., Consalvo, D., van Bladel, J.W.I.,
  van der Avoird, A., \& Pavlov-Verevkin, V. 1990, CP, 144, 81
\bibitem[Engeln \& Reuss (1991)]{enge91}
  Engeln, R., \& Reuss, J. 1991, CP, 156, 215
\bibitem[Fantoni et al. (1986)]{fant86} 
  Fantoni, R., van Helvoort, K., Knippers, W., \& Reuss, J. 1986, CP,
  110, 1
\bibitem[Fateley \& Miller(1962)]{fateley1962} Fateley, W.G., \&
  Miller, F.A. 1962, AcSpe, 18, 977 
\bibitem[Favre et al.(2014)]{favre2014} Favre, C., Carvajal, M.,
  Field, D., J\o rgensen, J.K., Bisschop, S.E., Brouillet, N.,
  Despois, D., Baudry, A., Kleiner, I., Bergin, E.A., Crockett, N.R.,
  Neill, J.L., Margul\'es, L., Huet, T.R., \& Demaison, J. 2014,
  \apjs, 215, 25
\bibitem[Fern\'andez-S\'anchez et al. (1989)]{jmf89b}
  Fern\'andez-S\'anchez, J.M., Valdenebro, A.G., \& Montero, S. 1989,
  JChPh, 91, 3327
\bibitem[Frisch et al. (2009)]{Gaussian} Frisch, M.J., Trucks, G.W.,
  Schlegel, H.B., Scuseria, G.E., Robb, M.A., Cheeseman, J.R.,
  Scalmani, G., Barone, V., Mennucci, B., Petersson, G.A. et al. 2009,
  Gaussian 09, Revision A.1, Gaussian, Inc., Wallingford CT
\bibitem[Groner \& Durig(1977)]{groner1977} Groner, P., \& Durig, J.R. 1977,
  JChPh, 66, 1856 
\bibitem[Groner et al.(1998)]{groner1998} Groner, P., Albert, S.,
  Herbst, E., \& de Lucia, F.C. 1998, \apj, 500, 1059 
\bibitem[Kasai \& Myers(1959)]{kasai1959} Kasai, P.H., \& Myers,
  R.J. 1959, JChPh, 30, 1096 
\bibitem[Koerber et al.(2013)]{koerber2013} Koerber, M., Bisschop,
  S.E., Endres, C.P., Kleshcheva, M., Pohl, R.W.H., Klein, A., Lewen,
  F., \& Schlemmer, S. 2013, A\&A, 558, 112
\bibitem[Kutzer et al.(2016)]{kutzer2016} Kutzer, P., Weismann, D., Wassmuth,
  B., Pirali, O., Roy, P., Yamada, K.M.T., Giesen, T.F 2016
  JMoSp 329, 28
\bibitem[Lovas et al.(1979)]{lovas1979} Lovas, F.J., Lutz, H., \&
  Dreizler, H. 1979, JPCRD, 8, 1051 
\bibitem[Neustock et al.(1990)]{neustock1990} Neustock, W.,
  Guarnieri, A., Demaison, J., \& Wlodarzak, G. 1990,
  ZNatA, 45, 702 
\bibitem[Niide \& Hayashi(2003)]{niide2003} Niide, Y., \& Hayashi,
  M. 2003, JMoSp, 220, 65 
\bibitem[Pople et al.(1987)]{pople1987} Pople, J.A., Head-Gordon, M.,
  Raghavachari, K. 1987, JChPh, 87, 5968
\bibitem[Richard et al.(2013)]{richard2013} Richard, C., Margul\`es,
  L., Caux, E., Kahane, C., Ceccarelli, C., Guillemin, J.-C.,
  Motiyenko, R.A., Vastel, C., \& Groner, P. 2013, A\&A, 552, 117
\bibitem[Schilke et al.(2001)]{schilke2001} Schilke, P., Bendford,
  D.J., Hunter, T.R., Lis, D.C., \& Phillips, T.G. 2001, \apj, 132,
  281 
\bibitem[Scuseria \& Schaefer III(1989)]{scuseria1989} Scuseria, G.E.,
    \& Schaefer III, H.F. 1989, JChPh, 90, 3700
\bibitem[Senent et al.(1995a)]{senent1995a} Senent, M.L., Moule, D.C.,
  \& Smeyers, Y.G. 1995, CaJPh, 73, 425 
\bibitem[Senent et al.(1995b)]{senent1995b} Senent, M.L., Moule, D.C.,
  \& Smeyers, Y.G. 1995, JChPh, 102, 5952 
\bibitem[Senent(1998a)]{senent1998a} Senent, M.L. 1998a, JMoSp, 191, 265
\bibitem[Senent(1998b)]{senent1998b} Senent, M.L. 1998b, CPL, 296, 299
\bibitem[Senent(2001)]{senent2001} Senent, M.L. 2001, ENEDIM, A
  variational code for non-rigid molecules
\bibitem[Senent et al.(2012)]{senent2012} Senent, M.L.,
  Dom\'{\i}nguez-G\'omez, R., Carvajal, M., \& Villa, M. 2012,
  JPhChA, 116, 6901 
\bibitem[Smeyers et al.(1996)]{Smeyers1996} Smeyers, Y.G., Villa,M., Senent, M.L., 1996, JMoSp, 177, 66.
\bibitem[Snyder et al.(1974)]{snyder1974} Snyder, L.E., Buhl, D.,
  Schwartz, P.R., Clark, F.O., Johnson, D.R., Lovas, F.J., \& Giguere,
  P.T. 1974, \apj, 191, L79 
\bibitem[Taylor \& Vidale(1957)]{taylor1957} Taylor, R.C., \& Vidale,
  G.L. 1957, JChPh, 26, 122 
\bibitem[Villa et al.(2011)]{villa2011} Villa, M., Senent, M.L.,
  Dom\'{\i}nguez-G\'omez, R., \'Alvarez-Bajo, O., \& Carvajal,
  M. 2011, JPhChA, 115, 13573  
\bibitem[Villa et al.(2013)]{Villa2013} Villa, M., Senent, M.L.,
  Carvajal, M., 2013, PCCP, 15, 10258
\bibitem[Western (2016)]{western2016} Western, C.M. 2016,
  JQSRT, 186, 221
\bibitem[Woon \& Dunning(1993)]{woon1993} Woon, D.E., Dunning,
  T.H. Jr. 1993, JChPh, 98, 1358
\end{thebibliography}
\end{document}